# Investigation of intrinsic magnetodielectric effect in $La_2CoMnO_6$: Role of magnetic disorder

J. Krishna Murthy[a*], K. Devi Chandrasekhar[a, b], S. Murugavel[c] and A.Venimadhav[a]

[a]Cryogenic Engineering Centre, Indian Institute of Technology, Kharagpur-721302, India

[b]Department of Physics and Center for Nanoscience and Nanotechnology, National Sun Yat-Sen University, Kaohsiung 804, Taiwan

[c]Department of Physics and Astrophysics, University of Delhi, Delhi 110007, India

## Abstract

We present a large magnetodielectric (MD) effect of 65 % at 100 kHz with 5 T field in B-site ordered $La_2CoMnO_6$ (LCMO) polycrystalline sample. Frequency and temperature dependent impedance and dielectric studies under magnetic field divulge both intrinsic and extrinsic origins for the observed MD effect. The temperature dependent Raman spectroscopy measurement has shown spin-lattice coupling that supports the intrinsic origin of the observed large MD response in LCMO. Extrinsic contributions to MD response mainly originate from disorder and interface effects; here, we signify this by hole carrier (Sr) doping at the A-site of the ordered LCMO sample. The comparison study has disclosed that with the disorder, the intrinsic polarization due to asymmetric hopping decreases significantly, and the disorder induced transport dominates in both MD and magnetoresistance behaviour with close resemblance.

# I    Introduction

Multiferroic systems exhibit spontaneous magnetization and electrical polarization simultaneously in a single phase are attractive for their potential usage in magnetoelectric (ME) memory devices.[1-3] The coupling between dielectric and magnetic ordering usually known as the second-order ME effect can be evaluated by measuring the change in dielectric permittivity under external magnetic field called as magnetodielectric (MD) effect.[4] For device applications, a large intrinsic MD response in a wide temperature range is desired. Recently, large MD effect was observed in diverse magnetic oxide systems and various intrinsic mechanisms like spin current[5], spin-lattice coupling[6], magneto-exchange striction[7,8] and charge ordering[9] were proposed. On the other hand, Maxwell-Wagner (MW) space charge polarization combined with magnetoresistance (MR) extrinsically contribute to MD behaviour in most of the polycrystalline dielectric oxide materials.[10] Therefore, evaluation of intrinsic MD response necessitates the complimentary experimental tools.

La- based double perovskite systems with general formula $La_2B'B''O_6$ (B'=Co/Ni & B''=Mn) have received a significant research interest due to their large MD and MR properties below their magnetic ordering.[11-13] In these systems, the ferromagnetic (FM) property is governed by Goodenough-Kanamori 180º superexchange interactions among the *B*-site cations.[14] The functional properties are strongly coupled to the crystal structure and the B-site ordering. The ordering depends on the synthesis conditions, doping on La or B'/ B" sites and particle size.[11, 15-18] Recently, Chandrasekhar *et al.,* have reviewed the effect of dimensionality on the magnetic transition and MD behaviour.[17] Highly ordered $La_2NiMnO_6$ system showed negligible MD effect while a small amount of antisite disorder (ASD) has shown a large MD effect.[11, 18] Further, in $La_2NiMnO_6$, the replacement of La by smaller rare earth ions influences the magnetic transition and shows the negligible extrinsic MD effect.[19] On the other hand, the isostructural $La_2CoMnO_6$ (LCMO) system has shown a significant MD effect that was ascribed to the spin-lattice coupling.[13, 20] Replacing La by smaller rare-earth elements (i.e., Y and Lu) has established an E*- type (↑↑↓↓) magnetic order that breaks the spatial inversion symmetry to manifest spontaneous electric polarization;[8, 21] these systems shown small MD effect.

Recently, the intrinsic MD response was reported in thin films[13] and single crystalline forms of LCMO[20]. On the other hand, both intrinsic as well as extrinsic MD effects were reported in LCMO nanoparticles.[16] There is no systematic study of MD behaviour in ordered LCMO sample. In this study, we report a large value of MD ~65 % at 100 kHz for 5 T in ordered LCMO sample and to our knowledge this is the highest MD effect observed in double perovskite systems. In order to explore the extrinsic origin of MD, we have introduced spin disorder by Sr doping at La site.

## II  Experiment:

Polycrystalline $La_{2-x}Sr_xCoMnO_6$ (x=0 and 0.25) ceramic samples were prepared by Sol-gel based chemical method. We have taken stoichiometric amounts of $La_2O_3$ (dissolved in $HNO_3$), $Sr(NO_3)_2.6H_2O$, $Co(NO_3)_2.6H_2O$, and $Mn(CH_3COO)_2$ in distilled water. Then ethylene glycol (EG) was mixed to the combined solution in such a way that the total volume of solution to EG ratio was maintained as 1:1.5. Further, a clear solution of EG complex metal nitrates was then evaporated by keeping the resultant solution on a hot plate at 100ºC for 30 minutes with continuous magnetic stirring followed by heating at 210ºC. During the evaporation, nitrate ions provide an in-situ oxidizing environment for EG, which partially converts the hydroxyl groups of EG to carboxylic acids. Towards the complete dehydration, the nitrates get decomposed leaving behind voluminous, carbonaceous, organic-based, black and fluffy powder which contains the desired metal ions in their matrix. The resultant precursor powder was sintered at 1300ºC for 24 hours to get the bulk polycrystals.

The crystallographic phase analyses of the samples were carried out using Phillips powder x-ray diffraction (XRD) system with Cu-$K_{\alpha 1}$ radiation. The surface morphology was characterized by JEOL-JSM5800 field emission scanning electron microscope (FESEM). Temperature and magnetic field dependence of DC and AC susceptibility measurements were done using Quantum Design SQUID-VSM magnetometer. Dielectric properties were performed using HIOKI 3532-50 LCR meter with an excitation AC voltage of 1 V by applying the silver paste contacts on both sides of the compacted pellet sample to make the parallel plate capacitor geometry. The temperature and magnetic field dependent dielectric and transport measurements were studied using a closed cycle-cryogen free superconducting magnet system from 300 K to 8 K. The surface of the sample was placed parallel to the direction of the applied magnetic field. The temperature dependent Raman spectra were recorded in the 180º backscattering geometry using a 514 nm excitation of air-cooled Argon Ion laser (Renishaw InVia Reflex Micro-Raman Spectrometer). Laser power at the sample was ~10 mW and typical spectral acquisition time was 2 min with the spectral resolution of 1 cm$^{-1}$. The temperature was controlled with an accuracy of ±0.1 K by using a Linkam THMS 600.

## III      Results and discussion:
**Structural characterization**

The XRD pattern of as-prepared LCMO sample has been verified with Rietveld refinement using FullProf Suite is shown in the Fig. 1(a). LCMO shows single phase and its XRD refinement fits well with the monoclinic crystal structure of P2$_1$/n space group. But the reflection corresponding to the double perovskite ordering of Co-Mn lattices could not be obtained from XRD as these transition metal ions have similar x-ray scattering factors. In the case of hole (Sr=0.25) doping, the structural refinement fitted well to the disordered rhombohedral crystal structure with R3c space group as shown in the Fig. 1(b). The structural information about the lattice parameters, bond length and bond angles were determined using Diamond software and are listed in Table. 1. As shown in the Table.1, hole-carrier doping increases bond angle between Co-O-Mn (~163.81°); which leads to more preferable FM alignment.[14] FESEM micrograph (see the inset of Fig. 1(a)) shows the average grains (Gs) size of the order of 3-5 $\mu$m and grain boundaries (GBs) size is <100 nm in the LCMO sample.

**Magnetic characterization**

Fig. 2(a) shows the temperature ($T$) variation of magnetization ($M$ ($T$)) with 0.01 T DC field for LCMO sample. A paramagnetic (PM) to FM ordering (T$_C$) has been observed at 230 K.[15] Temperature dependence of inverse magnetic susceptibility ($\chi^{-1}$) for 0.01 T in the PM region is fitted to the Curie-Weiss (CW) law (Fig. 2). From the fitted data, the effective PM moment ($\mu_{eff}$) and CW temperature ($\theta$) are found to be ~7.34 $\mu_B$ and ~215 K respectively, and the overall magnetic behaviour is consistent with the previous reports.[15] The $M$ ($H$) data measured at 5 K in LCMO sample (Fig. 2 (b)), shows a typical FM hysteresis loop with coercivity ($H_C$) ~0.5 T and saturation trend in magnetization at higher fields ($\geq$ 5 T). In the double perovskite systems, long range Co/Mn cation order is characterized by the saturation magnetization ($M_S$) estimated from the $M$ ($H$) loop. In the present work, the magnetization value at 5 K tends towards saturation limit at higher fields with the maximum magnetization value of ~5.75 $\mu_B$/f.u. at 7 T. While in the case of Sr doped sample, $M(H)$ loop at 5 K (as shown in the Fig. 2(b)) reveals a reduced $M_S$ of ~4.13 $\mu_B$/f.u. In case of ideal NaCl type of ordering at the B-site with $Co^{2+}$ and $Mn^{4+}$ ions, the total magnetization value would be ~6 $\mu_B$/f.u.[15] The presence of antisite defects and the additional magnetic disorder due to hole doping effectively reduces the magnetization. The Sr doping introduces $Co^{3+}$ at the cost of $Co^{2+}$ to maintain the charge balance similar to the Sr doped $La_2NiMnO_6$ polycrystalline samples.[22] This introduces additional spin disorder with various exchange interactions like; $Co^{2+}$-O-$Co^{3+}$, $Co^{3+}$-O-$Co^{3+}$, and $Co^{3+}$-O-$Mn^{4+}$. We have estimated the percentage of disorder using $M_S$= (1-2ASD) [$M_{Co}$+$M_{Mn}$] +$x$ (2ASD-1); where $M_{Co}$ and $M_{Mn}$ are the spin only magnetic moments of Co and Mn ions respectively and '$x$' denotes the amount of hole doping.[23] The first term in the above expression denotes the contribution of ASD disorder, while the second term indicates reduction in $M_S$ due to hole doping. Accordingly, we find 2 % and 7.5 % of ASD in pure and doped samples respectively.

In order to understand the magnetic behaviour of parent and doped samples, we have measured the temperature dependent in-phase component ($\chi'$) of AC susceptibility with an AC filed of ~1 Oe as shown in the Fig. 2(c) and 2(d). The parent compound shows the frequency independent single peak at ($T_C$) ~ 230 K that signifies the FM behaviour. The observed FM transition can be assigned to the superexchange interaction between $Co^{2+}$-$O^{2-}$-$Mn^{4+}$ and it is consistent with the previous reports.[13, 15] In the doped sample, an additional frequency independent magnetic ordering was observed at ~152 K ($T_{C2}$) which corresponds to the FM superexchange interactions of $Co^{3+}$-$O^{2-}$-$Mn^{4+}$. At lower temperatures ~95-105 K, a frequency dependent peak ($T_f$) in $\chi'$ has been noticed and this peak shifts systematically to higher temperature with the increase of frequency suggesting a glassy nature. The relaxation time ($\tau$) has been analyzed using the critical slowing down power law given by $\tau = \tau_0 (\frac{T_f - T_g}{T_g})^{-ZV}$ as shown in the inset of Fig. 2(d), where $\tau_0$ is the microscopic spin relaxation time, $T_g$ is the glassy freezing temperature and $zv$ denotes the critical exponent. From the fitting, the obtained values are; $\tau_0$ =4.54x10$^{-7}$ sec, $T_g$ =94.7 K and $zv$ =4.94. The large values of $\tau_0$ and $zv$ indicates the freezing of magnetic clusters rather than the individual atomic spins in doped sample.[24] The presence of various competing magnetic exchange interactions ($Co^{2+}$-$O^{2-}$-$Mn^{4+}$(FM), $Co^{3+}$-$O^{2-}$-$Mn^{4+}$(FM), $Co^{3+}$-$O^{2-}$-$Co^{3+}$(AFM), and $Co^{2+}$-$O^{2-}$-$Co^{2+}$(AFM)) along with ASD drives the the doped system to glassy state.[25]

**Magnetoresistance characterization**

Fig. 3(a) shows the temperature dependent DC electrical resistivity ($\rho$) under zero fields for ordered and disordered samples. Here, both the samples have shown the semiconducting behaviour with decreasing temperature and exhibit no anomaly near to magnetic ordering. The value of $\rho$ for doped sample is lower by an order of magnitude than the ordered sample at all temperatures, which indicates that the Sr doping enhances electrical conductivity in LCMO. The ln$\rho$ vs. $T$ data for both the samples fit well to the variable-range- hopping (VRH) mechanism,[26] $\rho = \rho_0 \exp[(\frac{T_0}{T})^{\frac{1}{4}}]$ within the measured temperature range as shown in the inset of Fig. 3(a). Here, $\rho_0$ and $T_o$ are constants; $T_o$ is given by $24/[k_B N(E_F)\xi^3]$, $N(E_F)$ is the density of localized states at the fermi level and $\xi$ is the decay length of the localized wave function (assumed to be average distance between two Co or Mn neighbor ions; ~0.396 nm for Sr=0 sample and ~0.390 nm for Sr=0.25 samples, taken from the structural refinement). From the fitted data the estimated values of $N(E_F)$ ~2.7x10$^{18}$/m$^3$ for Sr=0 and ~4.8x10$^{18}$/m$^3$ for Sr=0.25 indicates the density of states near to the Fermi level in the hole-doped LCMO sample is slightly higher than that of the ordered sample. Further, polaron activation energy ($W$) can be calculated using $T_0$ value as $W=0.25 k_B T_o^{0.25} T^{0.75}$; the estimated $W$ value varies between 0.11 eV to 0.23 eV (110-300 K) and 0.097 eV to 0.20 eV (100 -300 K) for Sr=0 and 0.25 samples respectively.

Fig. 3 (b) shows the temperature dependent MR (%) $= (\frac{R_{(H=5T)} - R_{(H=0T)}}{R_{(H=0T)}}) \times 100$ with 5 T field for both the samples. Though the temperature dependent resistivity data does not exhibit any anomaly near to magnetic ordering, the MR with a negative sign is apparent below FM ordering, and this indicates a strong correlation between transport and magnetic ordering. The negative MR increases linearly with the decrease of temperature in both the samples and it indicates the spin-dependent disorder scattering mechanism.[27] The isothermal field variation of MR at 140 K in Sr=0 and 0.25 doped samples is shown in the inset of Fig. 3(b). From the transport and MR measurements, it is clear that both electrical conductivity and MR values are higher in doped sample than the ordered compound. Such a high electrical conductivity, negative MR and the downward curvature of MR behaviour with field in doped sample suggests the double exchange mechanism is likely at play.[27, 28]

**Dielectric and magneto-dielectric behaviour**

Temperature dependent real part of dielectric permittivity ($\varepsilon'(T)$) and loss tangent factor (tan $\delta(T)$) for the ordered sample at different frequencies is shown in Fig. 4(a) & (b) respectively. From the dielectric data, one can notice the following features: (i) frequency dependent large value of $\varepsilon'$ ($\geq 10^4$) in the temperature window of 250-300 K, (ii) below 250 K a fall in $\varepsilon'$ by two orders of magnitude at characteristic temperature which depends on the measured frequency, and (iii) at low temperature (< 50 K), $\varepsilon'$ displays a frequency independent value of ~26. This value is higher than the isostructural La$_2$NiMnO$_6$ bulk system ($\varepsilon'$ ~10-15)[18] and LCMO nanoparticles ($\varepsilon'$ <10).[16] Such a high value of $\varepsilon'$ at low temperature suggests that the observed property in ordered LCMO bulk sample is intrinsic to the sample. However, the giant value of $\varepsilon'$ and its frequency dependent behaviour at high temperatures indicates that the dielectric property is strongly influenced by its microstructure and can be explained using boundary layer capacitance model;[29] where the large size Gs (3-5 μm) (as shown in the inset to Fig. 1(a)) are separated by the thin GB regions (<100 nm). Further, the tan $\delta(T)$ of the dielectric relaxation spectrum exhibits a broad peak (Fig. 4(b)) corresponding to a sharp drop in $\varepsilon'$(T), and the peak position shifts to higher temperature side with the increase of frequency. The relaxation time ($\tau$) can be expressed as,

$$\tau = \tau_0 \exp(\frac{E_a}{K_B T}) \quad \text{----------- (1)}$$

where $E_a$ is the activation energy required for the relaxation process, and $\tau_0$ is pre-exponential factor. The linear variation of ln$\tau$ vs. 1/$T$ (Fig. 4(c)) shows a good fit to the thermally activated Arrhenius mechanism (Eqn. (1)) with $E_a$=0.11 eV and $\tau_0$=1.04 x10$^{-9}$ sec. Under a magnetic field of 5 T, these values become ~0.09 eV and 2.47x10$^{-9}$ sec respectively.

In order to understand the influence of magnetic disorder on dielectric properties, we have compared the temperature dependent $\varepsilon'(T)$ data of ordered sample with Sr doped disordered LCMO at 1 kHz and 500 kHz as shown in Fig. 4(d). In disordered sample, three major distinguishes can be found from the ordered sample: (i) the relaxation step becomes broad, (ii) the frequency independent low temperature intrinsic $\varepsilon'$ value is reduced to half (~12), and (iii) at high temperature and low frequency (1 kHz) the $\varepsilon'$ value increases almost twice whereas for high frequency (500 kHz) it decrease by four times. On the other hand, the doped sample exhibits two dielectric relaxations in the tan $\delta$ vs. $T$ data (as shown in the inset to Fig. 4(c)) associated with two drops in the dielectric permittivity spectrum. The low temperature relaxation was ascribed to bulk (or) grain contribution while the high temperature relaxation can be related to the extrinsic origin of MW interface space-charge polarization. Moreover, the low frequency dielectric loss in this doped sample is large compared to the parent compound. The observed changes in the dielectric properties can be explained based on the influence of disorder on the local potential fluctuations as well as MW interfacial polarization. More commonly, in double perovskites, the intrinsic dipolar polarization can arise from the asymmetric hopping mechanism.[11, 30] This implies that the hopping have a high probability in a specific direction and leads to finite charge-transferred state with $Co^{(2+\delta)}$ and $Mn^{(4-\delta)}$. In the doped sample, the contribution of polarization from the asymmetric hopping reduces significantly due to the introduction of hole carriers. Moreover, the presence of higher ASD also contributes to the reduction of intrinsic dielectric polarization. The enhancement of $\varepsilon'$ at high temperature and low frequencies can be related to the MW interfacial polarization; where the enhanced conductivity of hole-doped Gs and accumulated charges at the GBs result in large space charge polarization.

In order to understand the MD response, we have investigated the temperature dependent dielectric and impedance properties of the samples under different magnetic fields. Fig. 5(a) shows the temperature dependent MD (%) $(= \frac{\varepsilon'_{(H=5T)} - \varepsilon'_{(H=0T)}}{\varepsilon'_{(H=0T)}}) \times 100$ for different frequencies at 5 T field in the ordered sample. At low temperatures below 50 K, a small (< 1%) and frequency independent MD can be noted but the data appears to be noisy. Because, due to the freezing of dipoles leads to low capacitance value approaching the limit of resolution of the instrument that causes random error and it is distinctly visible after taking the relative difference for MD % calculation. With increasing temperature, MD becomes maximum (~ 100 %) for 1 kHz at ~95 K and then decreases gradually above this temperature. With the increase of frequency, the value of MD decreases. The observed MD ~65% for 100 kHz at 140 K is very large compared to other double perovskite systems.[11, 31, 32]

The MD peak shifts towards high temperature side with the increase of frequency. Such a temperature and frequency dependent MD are often related to extrinsic origin, i.e. the combined contribution of grain conduction and MW effects.[10] However, a large MD of 65% at high frequencies cannot be accounted only by the extrinsic effects. Fig. 5(b) & (c) shows the variation of MD and magnetodielectric loss (ML (%) $=(\frac{\tan\delta_{(H=5T)}-\tan\delta_{(H=0T)}}{\tan\delta_{(H=0T)}})\times 100$ as a function of magnetic field for different frequencies. At low frequencies the opposite sign of MD and MDL can be observed, while, for frequencies ≥100 kHz, both MD and MDL shows the positive sign similar to the $LaCo_{0.5}Mn_{0.5}O_3$ single crystals.[20] In fact, the relaxation due to asymmetric hopping of charge carriers is dominant at high frequencies and contributes to the MD.[11]

In order to understand the effect of magnetic disorder on the MD effect, we have compared the field variation of isothermal MD (Fig. 5(d)) and MR (inset to Fig. 3(b)) responses at 140 K in both the samples. Here, LCMO shows a giant value of MD ~65 %, while Sr doped sample exhibits ~40 % for 100 kHz at 5 T. In the case of ordered sample the observed MR (~35 %) value is half of the MD, while in the disordered doped sample the observed MD value is comparable to its MR. Here, the comparison of MD is justified as the relaxation peaks in both the samples are around the same temperature range. On the other hand, in the disordered sample similar magnitudes of MD (%) and MR/ML (%) indicates that the dominant extrinsic origin is MW effect combined with negative MR.

**Temperature and magnetic field dependent Impedance spectra analysis**

The impedance spectroscopy (IS) studies allow us to reveal the multiple relaxations in terms of resistance and capacitance of Gs and GBs. In this regard, we have analyzed the impedance data to separate the intrinsic and extrinsic contributions of MD. The measured impedance data are usually represented by the Nyquist plots with imaginary and real parts of the impedance (-$Z''$ vs.$Z'$). Fig. 6(a) shows the complex impedance ($Z^*$) plots at 120 K for different magnetic fields. Here, we have observed two relaxation curves; large semicircular arc at low frequencies is related to the GBs, while the high frequency small arc is related to the Gs (i.e., bulk). Such a bulk related high frequency arc is obscured due to the large difference in resistance between Gs and GBs ($R_G$ ~25 kΩ and $R_{GB}$ ~1.3 MΩ) at 120 K. With an external magnetic field, a change in impedance at the GBs is large while a noticeable change can be observed at the Gs. Consequently, magnetic field effect on $\varepsilon'$ with frequency shows a clear variation in Gs as well as in GBs (Fig. 6(b)). Such a systematic increase of $\varepsilon'$ value under magnetic field at Gs suggests that the observed MD at high frequencies ($\geq$ 100 kHz) is indeed intrinsic to the ordered sample. In a magnetically ordered system, the spin-spin correlation function is strongly influenced by the application of external magnetic field that enhances the asymmetric hopping and $\varepsilon'$ value which leads to a large positive MD effect. To further clarify the intrinsic MD to spin-lattice coupling in parent LCMO, we have performed temperature dependent Raman spectroscopy measurements as discussed below.

**Temperature dependent Raman spectra characterization:**
In figure 7(a), we present the Raman spectrum of LCMO recorded at 300 K, which shows the prominent peak at 646 cm$^{-1}$ corresponding to the symmetric stretching mode. Further, the observed Raman shift at 491 cm$^{-1}$ in the present investigation confirms the mixed character of both anti-stretching and bending mode vibrations of (Co/Mn)O$_6$ octahedra. In order to understand the spin-lattice coupling in the present study, we have recorded temperature dependent Raman spectra from 300 K down to 77 K as shown in Fig. 7(b). The important observations from the temperature dependent Raman spectra are: (i) intensity of phonon excitations increases progressively with the decrease of temperature, while the full width at half maxima (FWHM) of the peaks decreases, (ii) phonon stretching mode shows a clear temperature dependence and exhibits anomalous softening of phonon frequencies below the FM ordering, and (iii) the anti stretching mode is found to be temperature independent. To verify the anomalous behaviour of stretching mode with spin-phonon coupling[34, 35] we have plotted the temperature dependence of phonon linewidth as shown in the inset of Fig. 7(b). Here, phonon linewidth decreases with the decrease of temperature and noticed a of small change in the slope around the magnetic ordering (T$_C$) ~230 K which is the signature of spin-phonon coupling in ordered LCMO.[35]

Further, the softening of stretching mode with temperature can be explained well with the standard anharmonic model proposed by Balkanski.[33] Here, the energy of the phonons can be described with the following equation,

$$\omega_{anh}(\text{T}) = \omega_0 - C\left(1 + \frac{2}{e^{\frac{\hbar\omega_0}{2k_BT}} - 1}\right) \quad \text{------------ (2)}$$

where, $\omega_0$ and $C$ are the fitting parameters and $\hbar\omega_0$ is the phonon energy. Temperature dependent symmetrical stretching mode is plotted and fitted to the anharmonic model using Eqn. (2) as shown in the Fig. 7(c). Below $T_C$, the deviation in stretching mode from the standard anharmonic model indicates the existence of spin-lattice coupling and the observed softening of phonon frequency ~3 cm$^{-1}$ value at low temperature is comparable with the LCMO thin films.[36] In strongly correlated magnetic systems, below the magnetic ordering the spin-lattice coupling arises from phonon modulation of the spin-exchange integral which depends on the amplitude of spin-spin correlations $<S_i.S_j>$ ($S_i$ and $S_j$ be the localized spins at the i$^{th}$ and j$^{th}$ states respectively).[37, 38] Considering the mean-field model, the phonon renormalization function is proportional to the magnetization as follows,

$$\Delta\omega \approx M^2(T)/M^2_{max}(0) \quad\quad\quad\quad ------------ (3)$$

where, $M(T)$ and $M(0)$ are the magnetization values at $T(K)$ and 0 K respectively. As shown in the Fig. 7(d), the scaling law in between the phonon renormalization function ($\Delta\omega$) and square of the magnetization indicates that a spin-lattice coupling exists below $T_C$. The temperature dependence of phonon modes below the FM transition suggests that spin-lattice coupling is involved significantly in the observed MD effect.

## IV    Conclusions:

In summary, we have investigated the structural, MD and the MR behaviours of LCMO and Sr doped LCMO polycrystalline bulk samples. Structural refinement of crystallographic data reveals that the hole-carrier doping enhances Co-O-Mn bond angle with change in the crystal structure from monoclinic to disordered rhombohedra. A giant value of $\varepsilon'$ in electrically heterogeneous LCMO sample is attributed to the interface boundary layer capacitance model. A comparative study of the frequency and temperature dependent impedance/dielectric studies combined the MR divulge both intrinsic and extrinsic origins for the observed MD effect. The magnetic disorder strongly suppresses the intrinsic MD contribution. Further, temperature dependent Raman studies elucidate the spin-lattice anomaly below magnetic ordering, which indicates the possible origin of intrinsic MD in the ordered LCMO system.

**Acknowledgments:**

The authors acknowledge DST, FIST in Cryogenic Engineering Centre and IIT Kharagpur for the funding VSM SQUID magnetometer. Krishna, thanks CSIR-UGC, Delhi for senior research fellowship (SRF).

*Table 1*: The list of structural information for Sr=0 and Sr=0.25 doped LCMO bulk system from the Rietveld fitting with goodness of fit ($\chi^2$) values are 1.3 and 1.24 respectively.

| La$_{2-x}$Sr$_x$CoMnO$_6$ | Lattice parameters | | | | | |
|---|---|---|---|---|---|---|
| | a (Å) | b (Å) | c (Å) | α(deg) | β(deg) | γ (deg) |
| Sr=0.0 | 5.5223(2) | 5.4862(2) | 7.7730(3) | 90 | 90.04(2) | 90 |
| Sr=0.25 | 5.4969(5) | 5.4969(5) | 13.2516(4) | 90 | 90 | 120 |
| | Bond distance and bond angles | | | | | |
| | Mn-O$_1$ (Å) | Mn-O$_2$ (Å) | Mn-O$_3$ (Å) | Co-O$_1$ (Å) | Co-O$_2$ (Å) | Co-O$_3$ (Å) |
| Sr=0.0 | 1.9821(3) | 1.9733(6) | 1.9821(3) | 2.0153(1) | 1.9863(2) | 2.0146(2) |
| Sr=0.25 | 1.9527(5) | -------- | -------- | 1.9527(1) | -------- | -------- |
| | Mn-O$_1$-Co (deg) | | Mn-O$_2$-Co (deg) | | Mn-O$_3$-Co (deg) | |
| Sr=0.0 | 160.873(1) | | 162.114(1) | | 151.728(1) | |
| Sr=0.25 | 163.81(2) | | -------- | | ------- | |

**Figure captions:**

[1] XRD pattern with Rietveld refinement of (a) Sr=0 and (b) Sr=0.25 doped LCMO samples; inset to (a) shows the FESEM micrograph of LCMO with an indication of Gs and GBs size.

[2] (a) Temperature dependent *M* (*T*) and $\chi^{-1}(T)$ data under 0.01 T, solid line indicates the CW law fit to the FC data for ordered LCMO sample, (b) isothermal field dependent *M* (*H*) loops at 5 K within ± 6 T fields for both Sr=0 and 0.25 doped LCMO samples, (c) and (d) shows the temperature dependent $\chi'(T)$ at different frequencies for Sr=0 and Sr=0.25 doped LCMO samples respectively, and inset to (d) shows the power law fit of τ vs. T$_f$ data.

[3] (a) Temperature dependence of $\rho$ under zero magnetic fields; inset shows the VRH fit to $\ln\rho$ vs. $T^{-0.25}$ relation, and (b) the temperature dependent MR (%) under 5 T; inset shows the isothermal field variationof MR (%) at 140 K for Sr=0 and 0.25 doped LCMO samples.

[4] Temperature dependence of, (a) $\varepsilon'(T)$ and (b) $\tan\delta\,(T)$ of LCMO ordered sample for different frequencies, (c) linear variation of $\ln\tau$ vs. 1/T with Arrhenius fit by using Eqn. (1) for LCMO sample, and (d) temperature dependent $\varepsilon'$ at 1 kHz and 500 kHz for both ordered and Sr doped LCMO samples, and inset shows the temperature variation of $\tan\delta\,(T)$ for three different frequencies (i.e., 1, 10 and 100 kHz) in Sr=0.25 doped sample.

[5] (a) Temperature dependent MD effect of LCMO with 5 T, isothermal field variation of (b) MD (%), and (c) ML (%) response at different frequencies for LCMO sample, and (d) shows comparison plot of isothermal field dependent MD (%) at 140 K and 100 kHz for Sr=0 and Sr=0.25 doped LCMO samples.

[6] (a) Isothermal Nyquist plots: $Z'$ vs. $-Z''$ at 120 K for different magnetic fields, and (b) frequency variation of dielectric permittivity ($\varepsilon'$) at different magnetic fields at 120 K for LCMO sample.

[7] Raman spectra of LCMO sample (a) at 300 K, and (b) at different temperatures, and inset to (b) shows the temperature dependent line width of the stretching mode (solid lines are guides to the eye), (c) temperature dependence of the relative shift in the stretching mode frequency from the anharmonic model and a solid line represents the fitting with Eqn. (2), and (d) temperature variation of the scaling law in between $\Delta\omega\,(T)$ and $[M\,(T)/M_s\,(0)]^2$ for stretching mode.

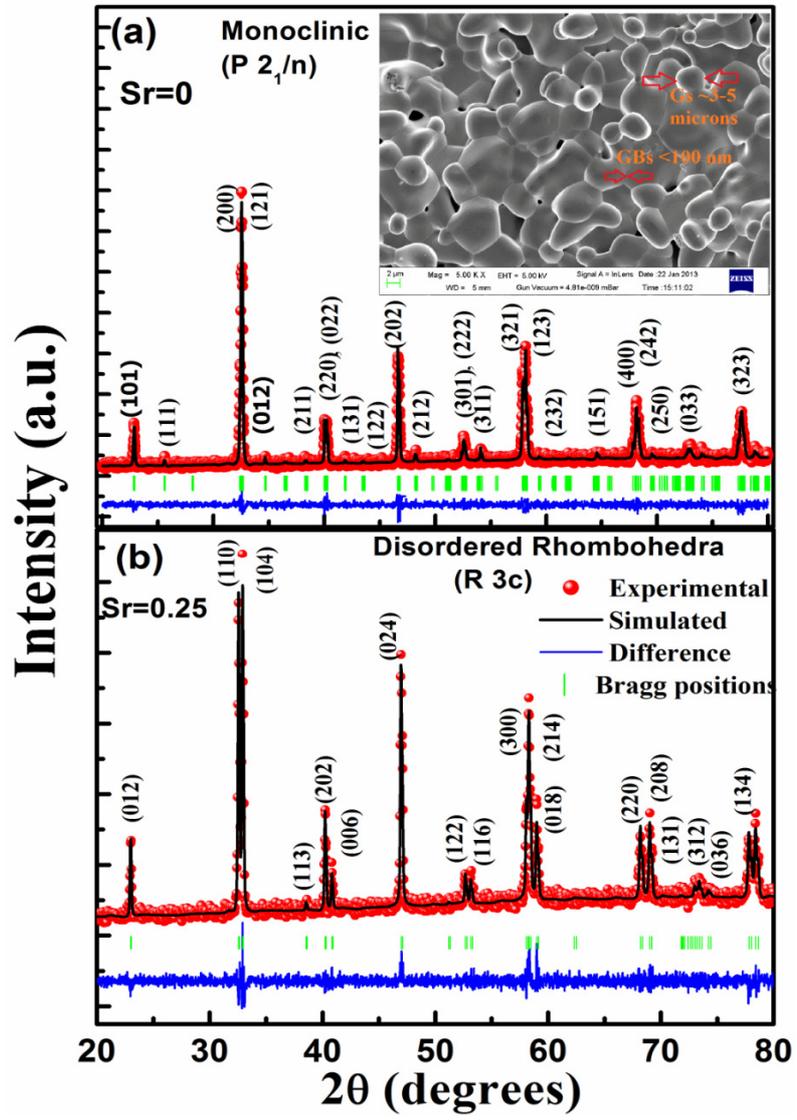

**Fig. 1**

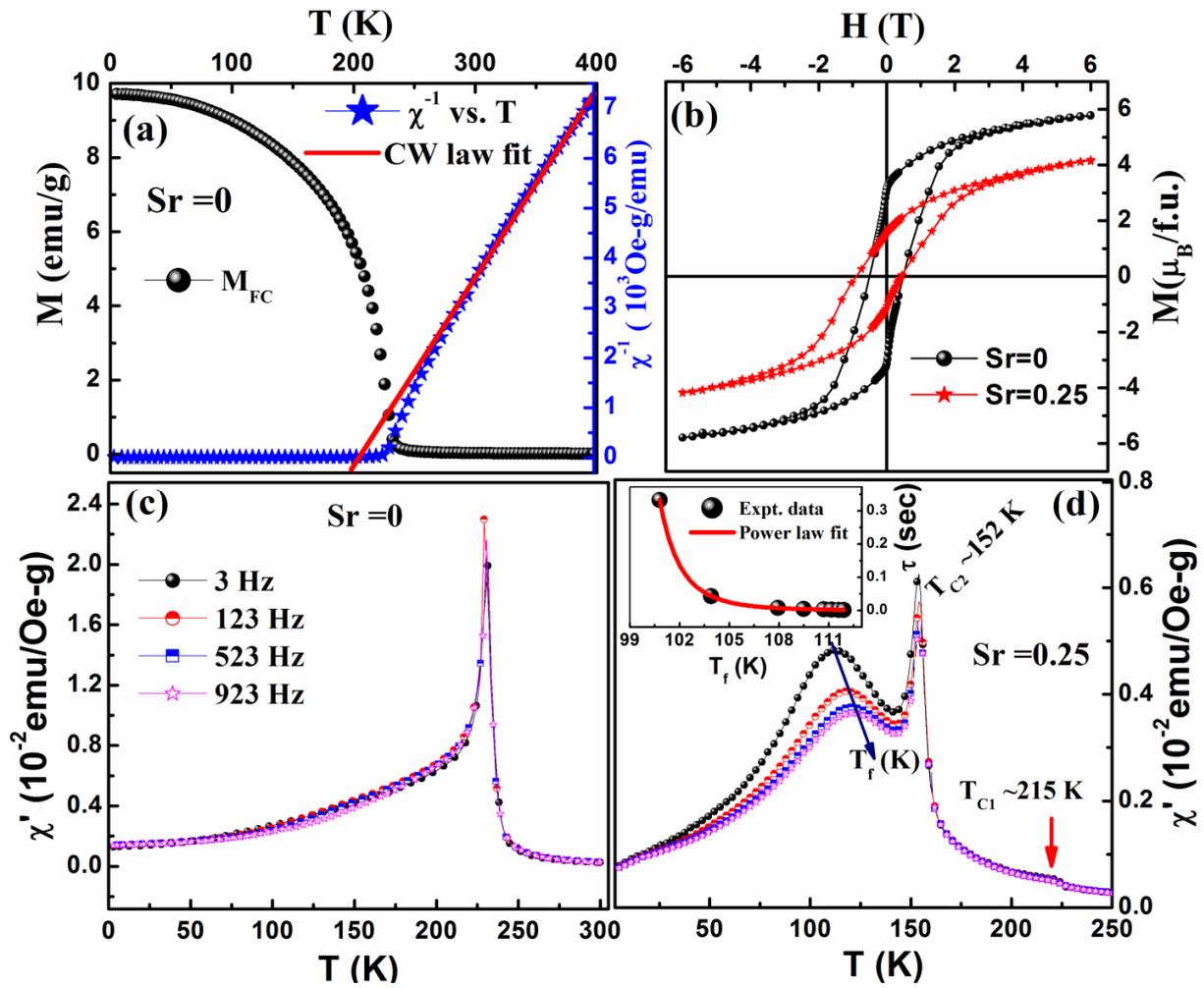

**Fig. 2**

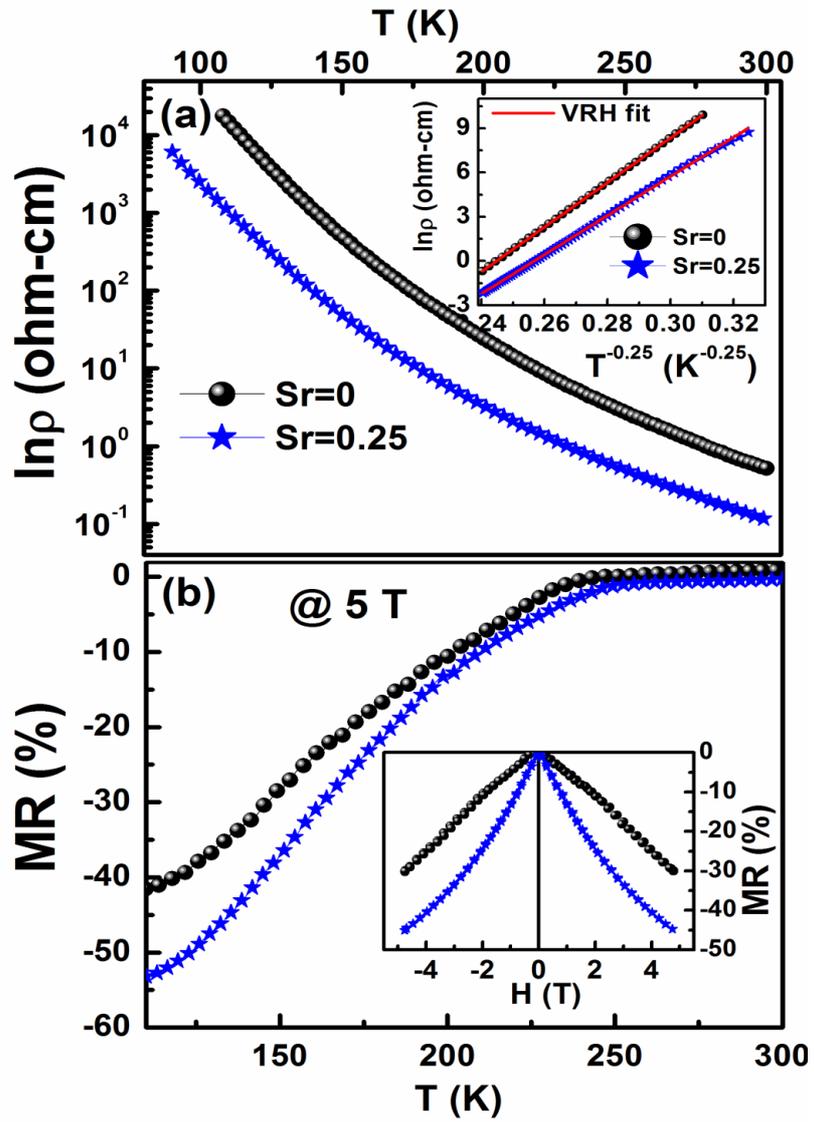

**Fig. 3**

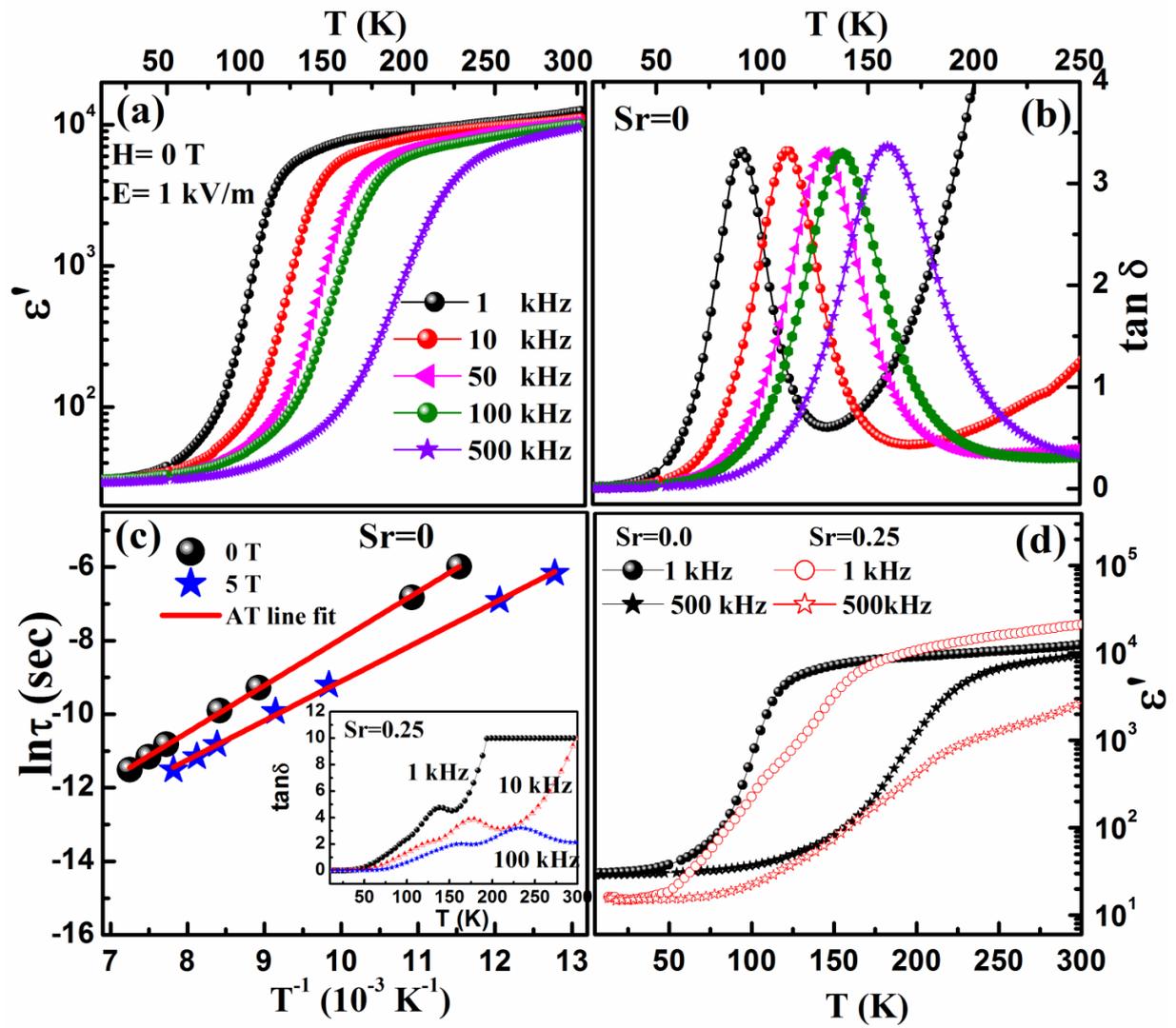

Fig. 4

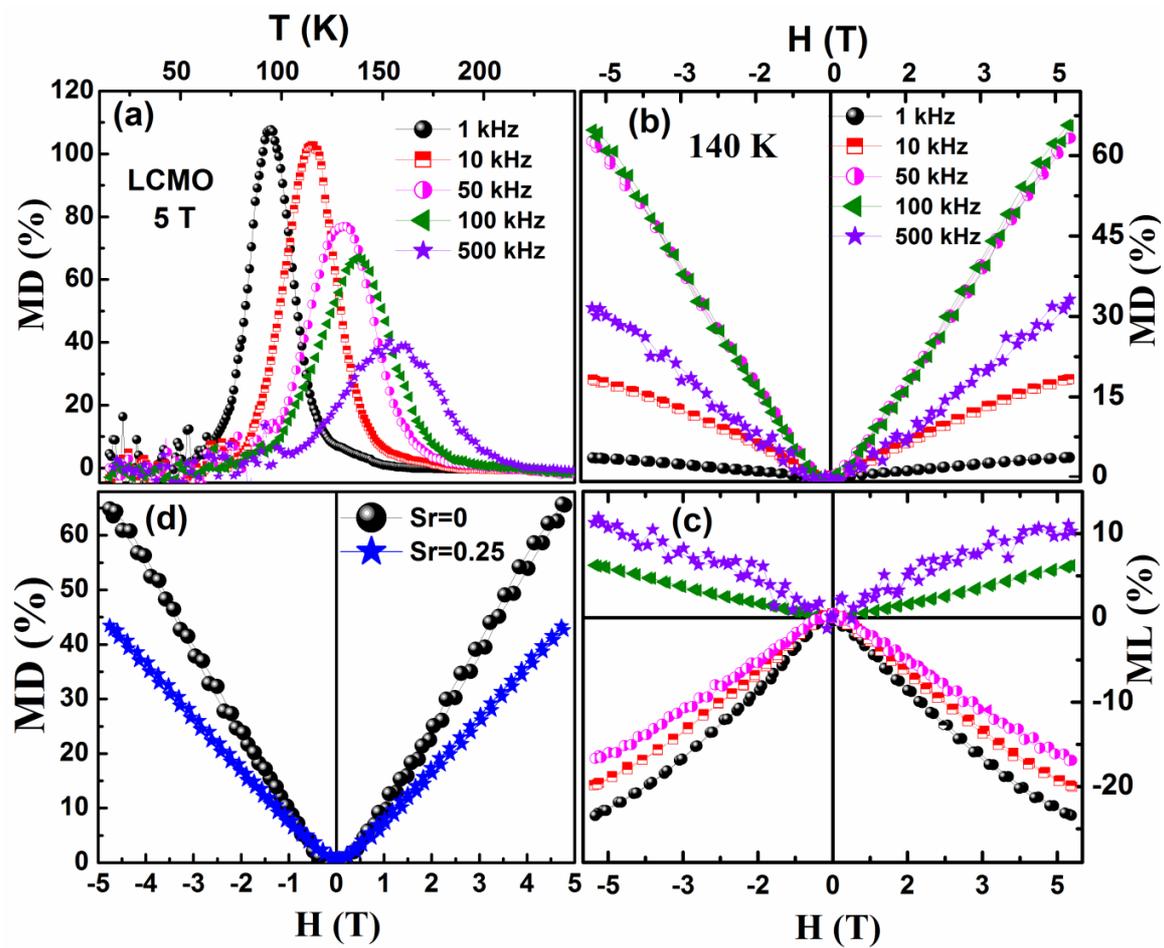

**Fig. 5**

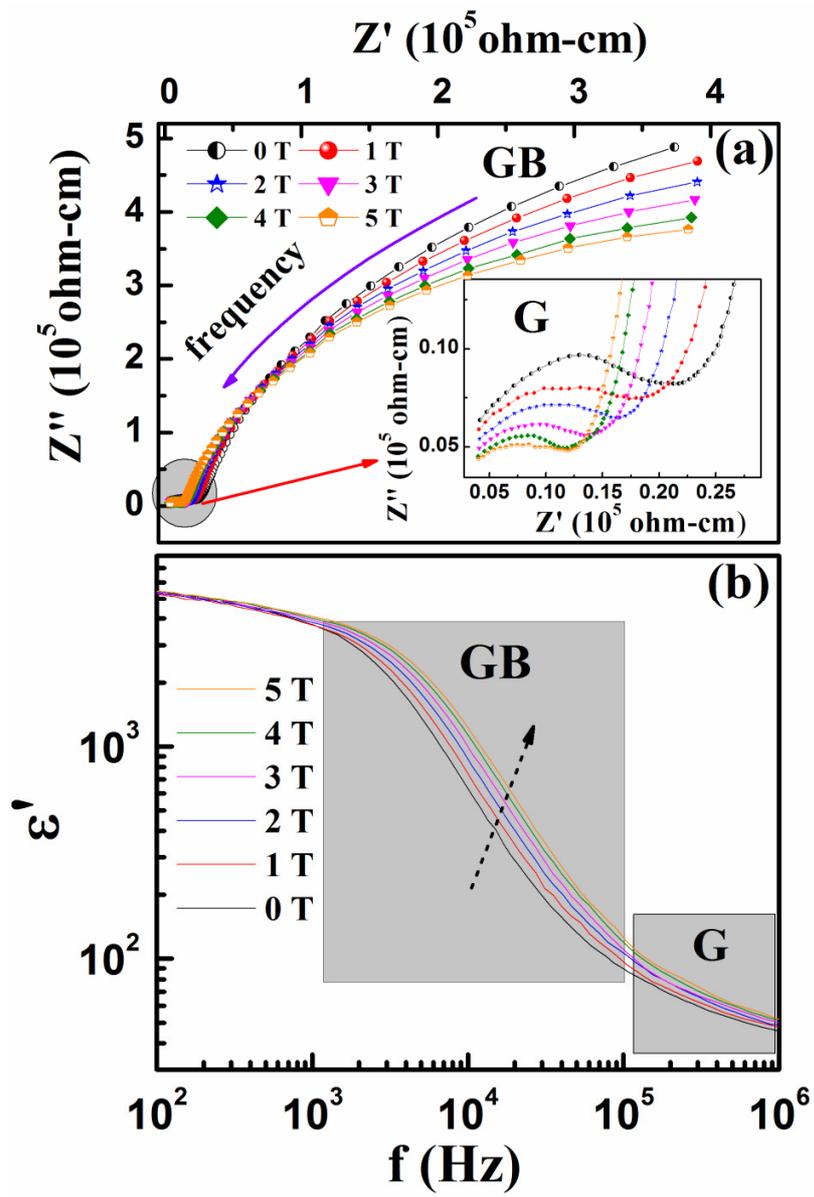

**Fig. 6**

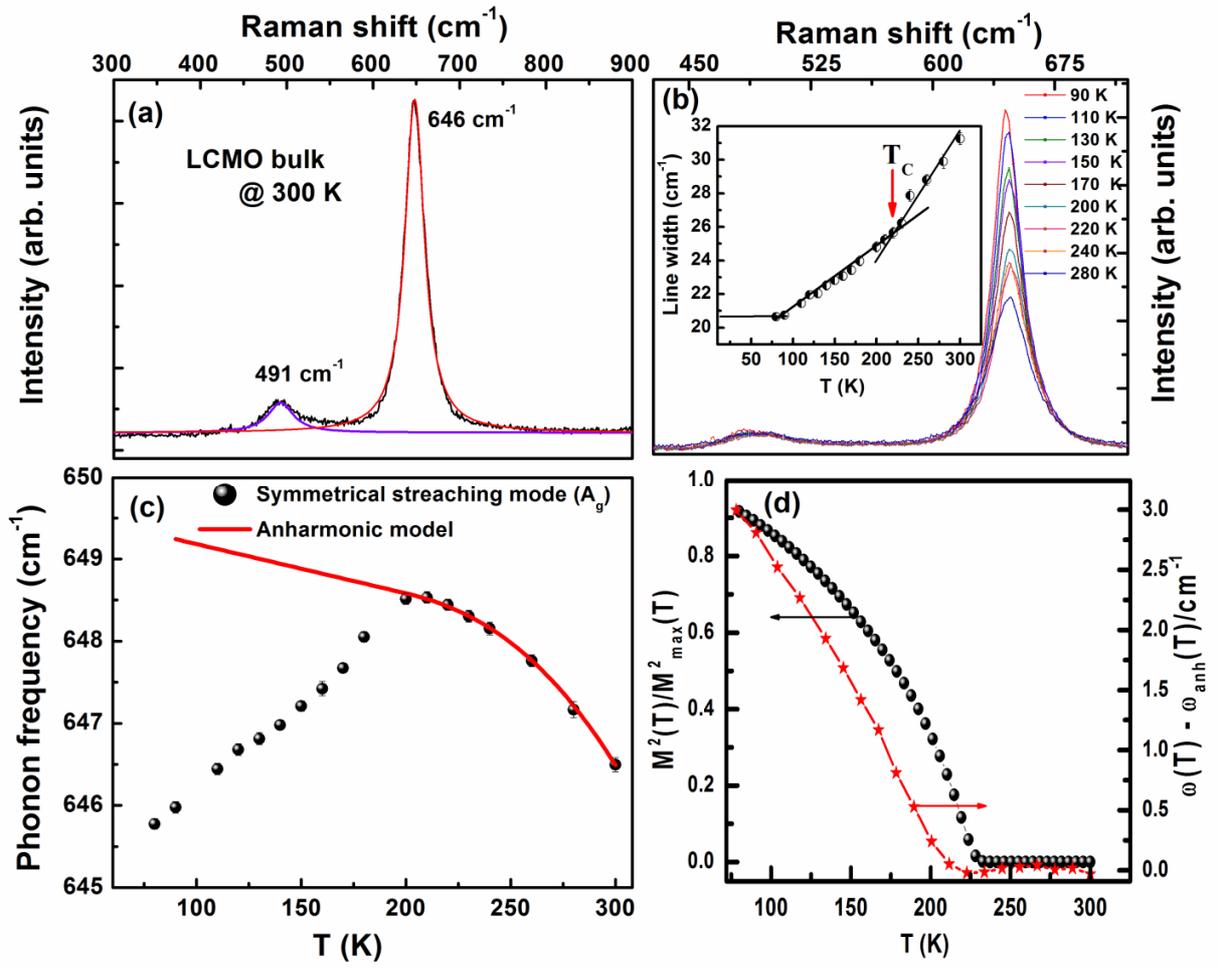

**Fig. 7**